\documentclass[12pt]{iopart}
\usepackage{graphicx}
\usepackage[table,xcdraw]{xcolor}
\usepackage{subcaption}

\usepackage{xcolor}
\colorlet{todocolor}{red}

\newenvironment{keywords}{\noindent\textbf{Keywords:} }{}
\begin{document}
\newcommand{\aperture}{Adaptive Aperture\textsuperscript{\texttrademark}}
\newcommand{\mevion}{Mevion S250i Hyperscan\textsuperscript{\texttrademark}}

\title[Interplay evaluation methodology for lung and esophageal cancer patients]{In vivo and predictive interplay evaluation methodology for lung and esophageal cancer patients treated in free breathing with IMPT}

\author{Giorgio Cartechini$^{1}$, Esther Kneepkens$^{1}$, Gloria Vilches-Freixas$^{1}$, Indra Lubken$^{1}$, Marije Velders$^{1}$, Sebastiaan Nijsten$^{1}$, Mirko Unipan$^{1}$, and Ilaria Rinaldi$^{1}$}
\address{$^1$ Maastricht University Medical Centre+, Department of Radiation Oncology (Maastro), GROW School for Oncology and Reproduction, Maastricht, The Netherlands}
%\ead{giorgio.cartechini@maastro.nl}

\begin{abstract}

\noindent
\textbf{Purpose:} Pencil beam scanning proton therapy is sensitive to respiratory motion, leading to potential dose inhomogeneities due to interplay effects. We developed and validated a predictive framework to assess interplay and motion robustness in lung and esophageal cancer patients treated under free-breathing conditions.

\noindent
\textbf{Methods:} A synthetic-breathing-based predictive model was implemented in the RayStation treatment planning system (TPS) and validated against an \textit{in vivo} approach using patient-specific respiratory traces and machine log files. Both were benchmarked against the Monte Carlo engine FRED. To demonstrate the methodology, the framework was applied to two clinical cases treated on the Mevion S250i system without rescanning. Dose accumulation incorporated respiratory phase, range ($\pm$3\%), and setup ($\pm$5~mm) uncertainties.

\noindent
\textbf{Results:} Excellent agreement was observed between TPS and FRED ($<$1\% mean dose difference) and between predictive and \textit{in vivo} models ($<2$\% across DVH metrics). Cumulative dose distributions for the primary CTV converged after five fractions, confirming robust delivery. 

\noindent
\textbf{Conclusion:} This automated, clinically integrated framework enables pre-treatment prediction and \textit{in vivo} validation of interplay and motion robustness in PBS plans. Preliminary results support its clinical utility, especially for hypofractionation and targets with large motion ($>$2~cm). A larger cohort study is ongoing and will be reported separately.
\end{abstract}


%\submitto{\PMB}
\maketitle

\section{Introduction}
Pencil beam scanning (PBS) proton therapy, particularly in the form of intensity-modulated proton therapy (IMPT), is the most advanced proton delivery technique, offering excellent dose conformity and enhanced sparing of surrounding healthy tissues~\cite{chang2017consensus}. However, its layer-by-layer, spot-scanning delivery is inherently sensitive to respiratory motion, particularly in thoracic and upper abdominal tumors~\cite{chang2014clinical}. This motion sensitivity can lead to the so-called interplay effect, a temporal mismatch between tumor motion and beam delivery, resulting in dose inhomogeneities such as cold and hot spots within the target volume~\cite{bert2011motion}.

Several studies have investigated the magnitude and clinical implications of this effect. Li et al.~\cite{li2014interplay} quantified interplay-induced deviations in stage III lung cancer patients, showing that although fractionation mitigates some heterogeneity, significant variations can still occur in individual fractions. Complementary, Monte Carlo simulations by Grassberger et al.~\cite{grassberger2015motion} further demonstrated that small spot sizes and rapid energy switching can exacerbate dose non-uniformity, particularly in single-fraction deliveries.

To mitigate the impact of respiratory motion, several strategies have been investigated, including 4D robust optimization~\cite{engwall20184d, spautz2023comparison, bengtsson2025interplay} and various techniques, including rescanning~\cite{rana2021investigating, knopf2011scanned}. Despite their effectiveness, these methods can be complex and time-consuming, limiting their routine clinical implementation. As a result, more streamlined and time-efficient approaches are often preferred. At our institution, patients with lung~\cite{taasti2025proton} and esophageal cancers ~\cite{canters2024robustness} are treated in free-breathing using 3D robust optimization based on an Internal Target Volume (ITV), defined as the union of Clinical Target Volume (CTV) contours across all respiratory phases. This ITV-based strategy, combined with planning on an average 4DCT and 3D and 4D robust evaluations~\cite{taasti2021treatment}, incorporates motion into the optimization while preserving clinical feasibility.

Treatments are delivered using the Mevion Hyperscan S250i system (Mevion Medical Systems, Littleton, MA, USA)~\cite{vilches2020beam}, which accelerates a fixed energy beam of around 230 MeV. The beam energy is modulated via eighteen range shifter plates within the nozzle. This configuration results in relatively large spot sizes, advantageous characteristics for mitigating the impact of motion in PBS delivery~\cite{rana2022small}.

The so-called
    `4D evaluation' framework implemented at our institution to assess plan robustness~\cite{taasti2021treatment} accounts for anatomical variations across the breathing cycle but does not incorporate the temporal relationship between beam delivery and respiratory motion. As a result, it does not fully capture the dosimetric consequences of the interplay effect. To address this limitation, we propose a comprehensive framework for evaluating both interplay effects and motion robustness in PBS plans for lung and esophageal cancer patients treated under free-breathing conditions.

We developed a predictive method to simulate interplay effects over the entire treatment course using synthetic respiratory scenarios, enabling evaluation to be performed prior to the start of treatment. The model is validated against an \textit{in vivo} approach that integrates fraction-specific respiratory traces and machine log files recorded during treatment. The evaluation framework is fully automated, implemented within the clinical treatment planning system, and further validated using the independent Monte Carlo dose engine FRED~\cite{schiavi2017fred}, clinically validated in our proton facility~\cite{gajewski2020implementation}. This article focuses on presenting the methodology for both predictive and \textit{in vivo} evaluations; extension of this work to a larger patient cohort is currently underway and will be the subject of a separate publication.

\section{Materials and Methods}
\subsection{Interplay effect evaluation: Predictive and \textit{in vivo} methods}
\label{sec:InterplayEval}

The interplay evaluation was conducted using two approaches: predictive and \textit{in vivo}. The predictive approach is designed to be performed before the start of the treatment, without prior knowledge of the patient’s breathing trace. In contrast, the \textit{in vivo} approach retrospectively evaluates plan robustness to organ motion by incorporating the patient-specific breathing traces per fraction acquired during treatment and the corresponding machine log files recorded after each fraction.

\begin{figure}
    \centering
    \includegraphics[width=\linewidth]{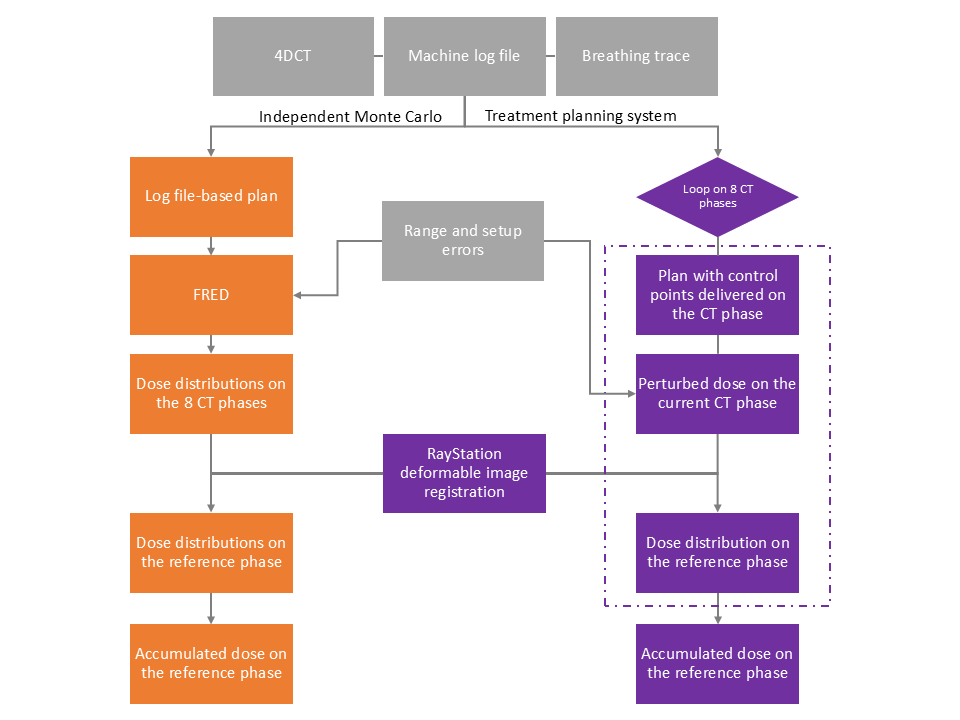}
    \caption{Workflow for the interplay evaluation using the FRED independent Monte Carlo engine (orange blocks) and the clinical TPS, RayStation (purple blocks). The dashed line groups the blocks within the loop over the 8 4DCT phases. The grey blocks represent the general input models independently of the dose engine used. 
    This workflow was applied to both the predictive and \textit{in vivo} approaches.}
    \label{fig:interplayFlowchart1}
\end{figure}

\textit{In vivo} and predictive interplay evaluations were performed using two Monte Carlo dose engines: FRED v3.76.4~\cite{schiavi2017fred}, and the clinical treatment planning system (TPS) RayStation (RaySearch Laboratories, Stockholm, Sweden) version 12B (Monte Carlo v5.4). The interplay workflow is illustrated in Figure~\ref{fig:interplayFlowchart1}. The 4DCT, breathing traces, and a machine log file are used as input for both Monte Carlo systems. The 4DCT represents the anatomical model of the patient’s geometry. It consists of eight 3D CTs corresponding to different respiratory phases:  0\%, 25\%, 50\%, 75\%, and 100\% inhale, and 25\%, 50\%, 75\% exhale. The ANZAI belt system (Anzai Medical Co. Ltd, Tokyo, Japan) was used to measure the patient’s respiratory amplitude during the CT scan. The 3D volumetric images were sorted based on the acquired breathing signals to assign them to the appropriate respiratory phases~\cite{lu2006comparison, li2006technical}.

The \textit{in vivo} approach exploits the fraction-specific breathing traces acquired by the C-RAD system \cite{walter2016evaluation} installed in the treatment room. Since the predictive approach is intended to be performed before treatment is delivered and patient-specific breathing traces are not available, we generated 24 synthetic breathing curves based on a sinusoidal function model, varying the breathing period (2 s, 3.5 s, and 5 s) and the starting phase according to the eight CT phases. By comparing predictive and \textit{in vivo} approaches, we will show that these 24 traces represent a range of plausible breathing scenarios for the patient.

To map each beam spot to one of the eight phases of the 4DCT, we used fraction-specific machine log files for the \textit{in vivo} model. For the predictive model, log files can be generated before the start of treatment through a dry run of the plan on a water phantom, similar to the patient quality assurance (QA) procedure.

Following the methodology described by Pastor-Serrano et al.~\cite{pastor2021should}, we combined 4DCT phases, patient-specific breathing traces, and machine log file data to assign each proton spot to a corresponding 4DCT phase. Dose was then recalculated with both FRED and TPS Monte Carlo on all eight 4DCT phases. Using RayStation’s deformable image registration (DIR) algorithm, the resulting dose distributions were warped onto a reference phase (50\% exhale) to compute the cumulative dose (Figure~\ref{fig:interplayFlowchart1}).

To assess the robustness of the plan against organ motion, as well as setup and range uncertainties, we systematically applied an isocenter shift of $\pm$5~mm along each Cartesian axis, combined with a systematic range error of $\pm$3\% in the CT calibration curve. To reduce computational time, we employed a probabilistic approach: for each breathing scenario, one random combination of range and setup error (representing a worst-case scenario) was sampled.

\begin{figure}
    \centering
    \includegraphics[width=\linewidth]{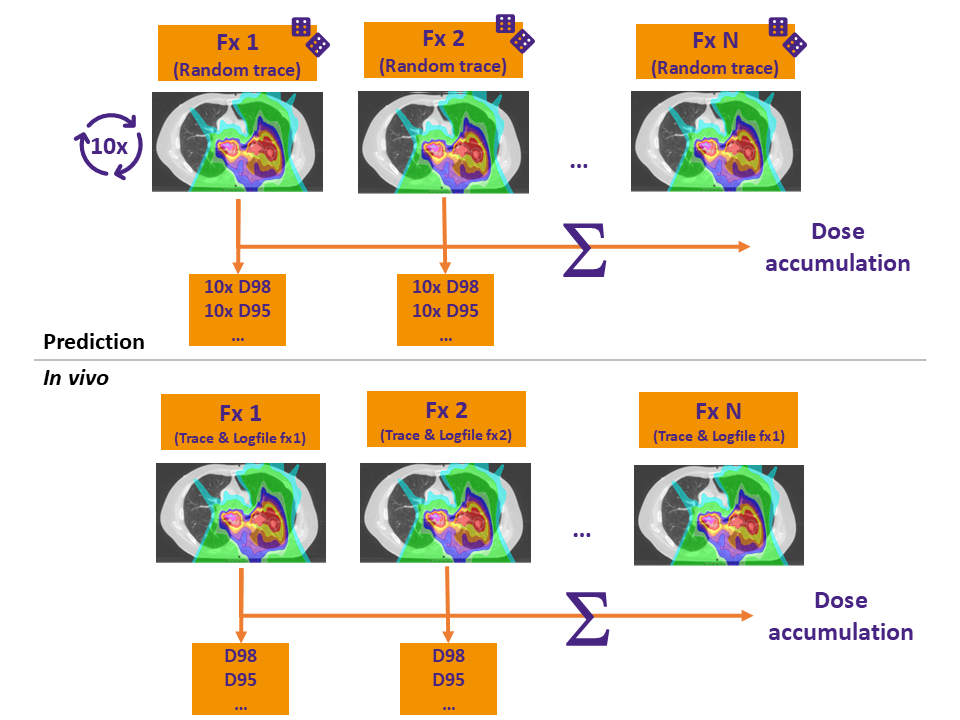}
    \caption{Top: Dose accumulation over treatment fractions for the predictive approach. A synthetic fractionation scheme was generated by randomly sampling a dose distribution scenario for each treatment fraction. This process was repeated ten times to assess variability. Bottom: Dose accumulation for the \textit{in vivo} approach using fraction-specific log files and breathing traces for each treatment session.}
    \label{fig:interplayFlowchart2}
\end{figure}

The effect of fractionation was quantified by accumulating the dose distribution over the entire treatment course, as illustrated in Figure~\ref{fig:interplayFlowchart2}. For the predictive method, a synthetic treatment course was simulated by randomly assigning one of the 24 breathing scenarios to each treatment fraction. The cumulative dose across all fractions was then calculated, and dose–volume histogram (DVH) parameters were extracted. Specifically, for each cumulative dose distribution, DVH metrics were computed for the primary tumor CTV (CTVp) and nodal CTV (CTVn), including D98 and D95 (minimum dose received by 98\% and 95\% of the volume, respectively) and V95 (volume receiving at least 95\% of the prescribed dose). To assess variability, the simulation was repeated ten times, and the mean and standard deviation of the DVH parameters are reported.

For the \textit{in vivo} evaluation, we recalculated the dose distribution from the fraction-specific breathing trace and machine log file, and then accumulated the doses over all fractions. As described previously, we extracted the DVH parameters D98, D95, and V95.

To quantitatively compare the inter-fractional distribution of DVH parameters obtained from the \textit{in vivo} evaluation and the predicted scenarios, we applied the statistical analysis method described by Pastor-Serrano et al.~\cite{pastor2021should}. For each quantity of interest $\Theta$ (i.e., D98, D95 and V95), we approximated the distribution across the 24 breathing trace scenarios (or treatment fraction for the \textit{in vivo}) using a percentile vector, $\delta_{\Theta} = \{median, 2, 98 \}$. We then computed the Relative Difference Error (RDE), defined as the average relative difference between two percentile vectors associated with different distributions of the same quantity of interest:
\begin{equation}
    RDE(\delta_{\Theta,1}, \delta_{\Theta,2}) = \frac{1}{n_i}\sum_{i}^{n_i}\frac{|\Theta_{i,1}-\Theta_{i,2}|}{\Theta_{i,2}} \times 100
\end{equation}
where $n_i$ is the number of elements in the percentile vector.

\subsection{FRED: Independent Monte Carlo engine}
Dose distributions were recalculated using the fast GPU-based Monte Carlo dose engine FRED~\cite{schiavi2017fred}, accounting for both organ motion and the temporal dynamics of beam delivery. The Mevion S250i beam model was implemented in FRED and clinically validated against experimental commissioning data to accurately reproduce the beam characteristics used at our institution~\cite{gajewski2020implementation}.

The 4DCT was imported into FRED by loading the eight 3D CT phases as standard geometry volumes. Efficient RAM and GPU memory management were required due to the large memory footprint of storing all CT phases simultaneously.

FRED supports dynamic activation and deactivation of geometry components during simulation. Leveraging this functionality, we activated the corresponding 4DCT phase during the delivery of each individual spot to model motion-resolved dose deposition. To account for setup and range uncertainties, we introduced perturbations in each simulation by modifying the Hounsfield Unit–Relative Stopping Power (HU-RSP) calibration curve by $\pm$3\% and by applying isocenter shifts of $\pm5$~mm in all directions. As a result, eight separate dose distributions were generated, each corresponding to one of the 4DCT phases.

The resulting dose distributions were imported into the RayStation TPS, where deformable image registration was applied to map each distribution onto a common reference phase. The final accumulated dose was then computed by summing the warped dose distributions.

\subsection{RayStation TPS Evaluation}
RayStation version 12B was used for dose recalculation, employing the Monte Carlo v5.4 algorithm with a final dose calculation uncertainty of 1\%. To streamline the workflow, we integrated the interplay evaluation directly into the clinical treatment planning system using a Python (v3.8) script via the RayStation API, automating the steps outlined in Figure~\ref{fig:interplayFlowchart1}. The evaluation followed the same principles as those implemented in FRED: Control points were mapped to the corresponding 4DCT phases, and dose distributions were recalculated for each phase. Range and setup uncertainties were introduced by perturbing the dose distributions accordingly. As in the FRED-based approach, RayStation’s DIR algorithm was used to map all eight phase-specific dose distributions onto a common reference phase, after which the cumulative dose was computed.

\subsection{Lung and Esophageal Cancer Patient Cases}

\begin{figure}
    \centering
    \includegraphics[width=\linewidth]{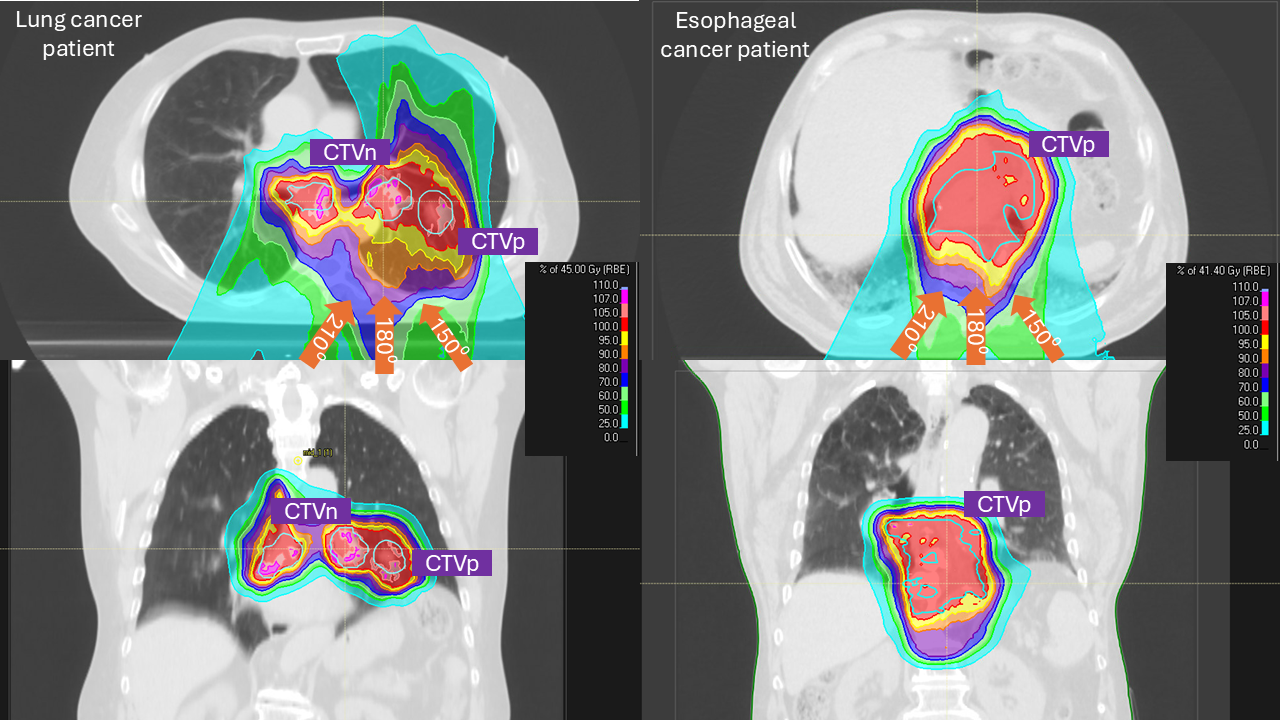}
    \caption{Axial and coronal view of the lung and esophageal cancer patients. Both cases were irradiated with three beams at 150$^\circ$, 180$^\circ$ and 210$^\circ$ gantry angles.}
    \label{fig:Patientsimages}
\end{figure}

We applied the proposed interplay evaluation methodology—comprising both predictive and \textit{in vivo} approaches—to two clinical cases, a lung and an esophageal cancer patient treated at our institution to test the methodology workflows and assess clinical feasibility (Figure~\ref{fig:Patientsimages}).

The lung cancer patient was characterized by a gross tumor volume (GTV) amplitude, defined as the maximum displacement of the GTV center of mass across 4DCT phases, of 9.7 mm. The treatment plan employed three posterior beams and was 3D robustly optimized on the ITV expanded by 1 mm to account for delineation uncertainties, incorporating 3\% range and 5 mm setup errors. Further details on the lung cancer patient’s plan setup and optimization are provided by Taasti et al.\cite{taasti2025proton}.

Similar to the lung case, the esophageal cancer patient was planned using three posterior beams and robustly optimized on the ITV expanded by 3 mm, incorporating 3\% range and 5~mm setup uncertainties. Additional details are provided by Canters et al.~\cite{canters2024robustness}.

\section{Results}

This section presents a systematic comparison between (i) the FRED Monte Carlo dose engine and the RayStation TPS for interplay evaluation, and (ii) the predictive and \textit{in vivo} methods for modeling respiratory motion. The comparison between FRED and RayStation TPS was conducted using a lung cancer case, while the analysis of predictive versus \textit{in vivo} approaches was performed for both lung and esophageal cancer patients. Unless otherwise stated, all results reported in Section~\ref{sec:ResInterplay} were generated using the FRED dose engine.

\subsection{Comparison between FRED and RayStation TPS}

\begin{figure}
    \centering
    \includegraphics[width=\linewidth]{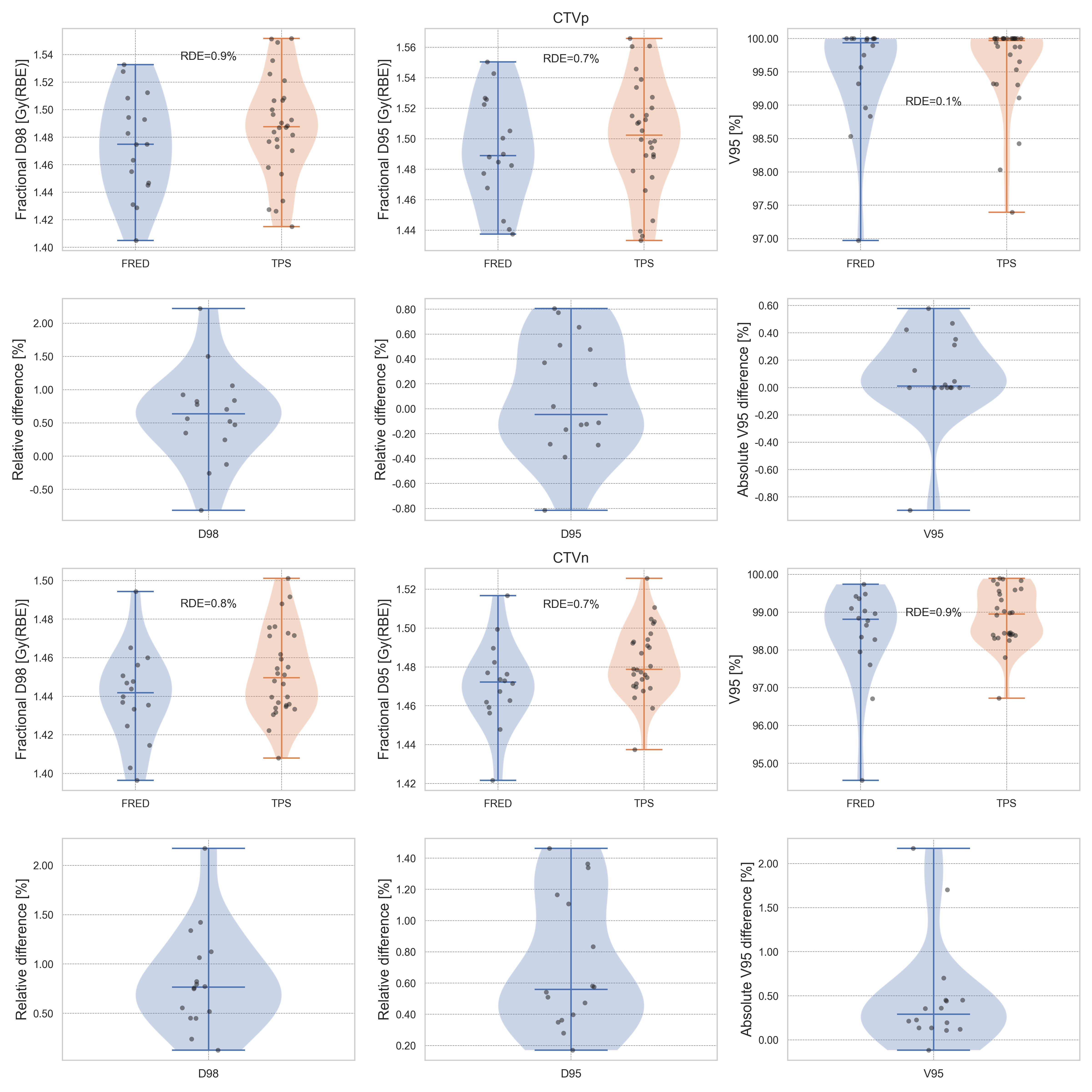}
    \caption{Comparison of D98, D95, and V95 between RayStation TPS and FRED dose engines using the \textit{in vivo} approach for the lung cancer case.}
    \label{fig:FREDVsTPS}
\end{figure}

To verify consistency between dose engines, we compared interplay-affected dose distributions computed with FRED and RayStation using the \textit{in vivo} method. As illustrated in Figure~\ref{fig:FREDVsTPS}, both engines yielded similar DVH parameter values for both the CTVp and CTVn. The median absolute difference between the two dose engines was below 1\% for all metrics, with only a single fraction showing a deviation above 2\% for D98. RDE values for D98, D95, and V95 remained consistently below 1\%, confirming a high level of agreement between the two dose calculation engines.

\subsection{Interplay Evaluation: Predictive vs. \textit{In Vivo} Approach}
\label{sec:ResInterplay}

\begin{figure}[htbp]
    \centering
    \begin{subfigure}[b]{0.8\textwidth}
        \centering
        \includegraphics[width=\linewidth]{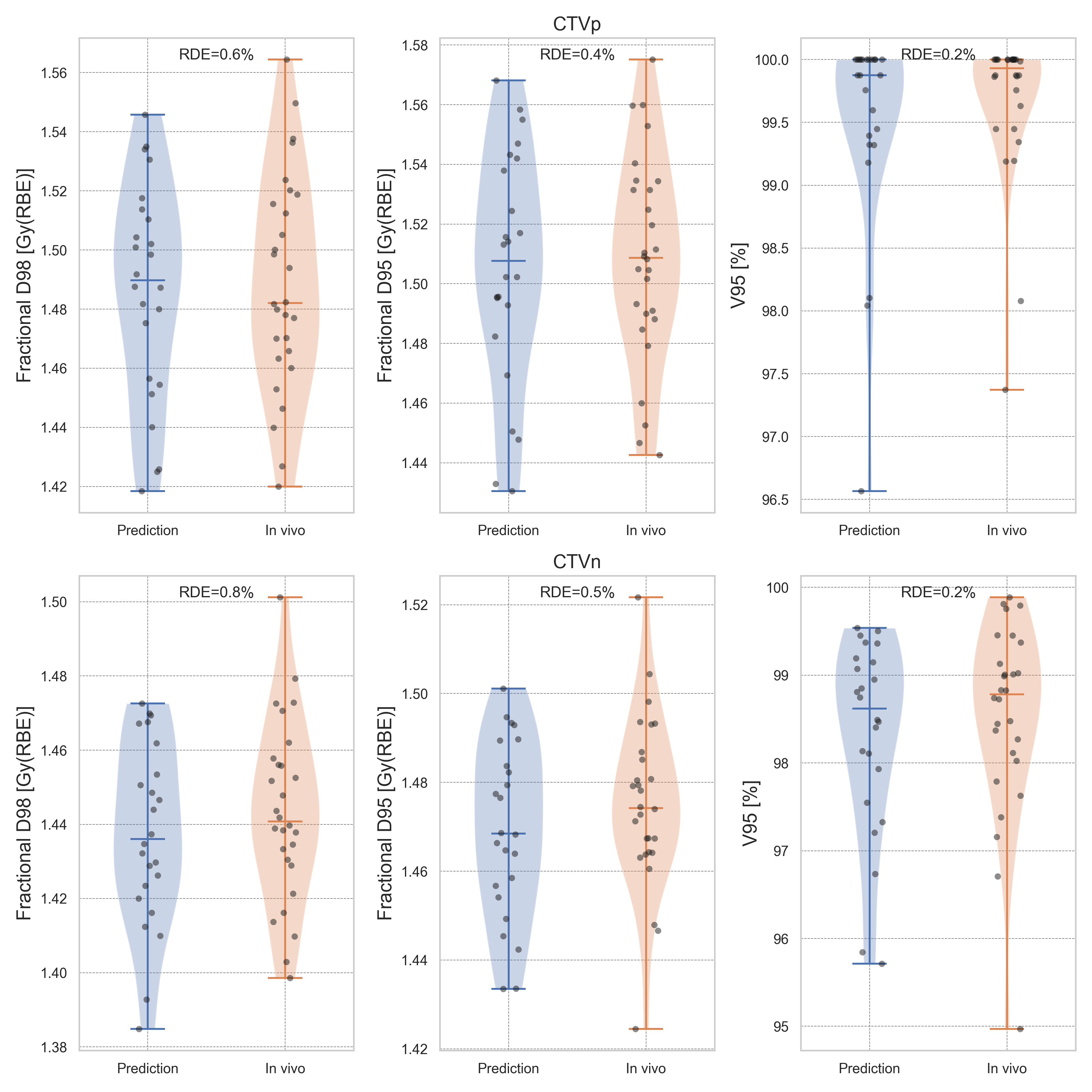}
        \caption{}
        \label{fig:subfig1}
    \end{subfigure}
    \hfill
    \begin{subfigure}[b]{0.8\textwidth}
        \centering
        \includegraphics[width=\linewidth]{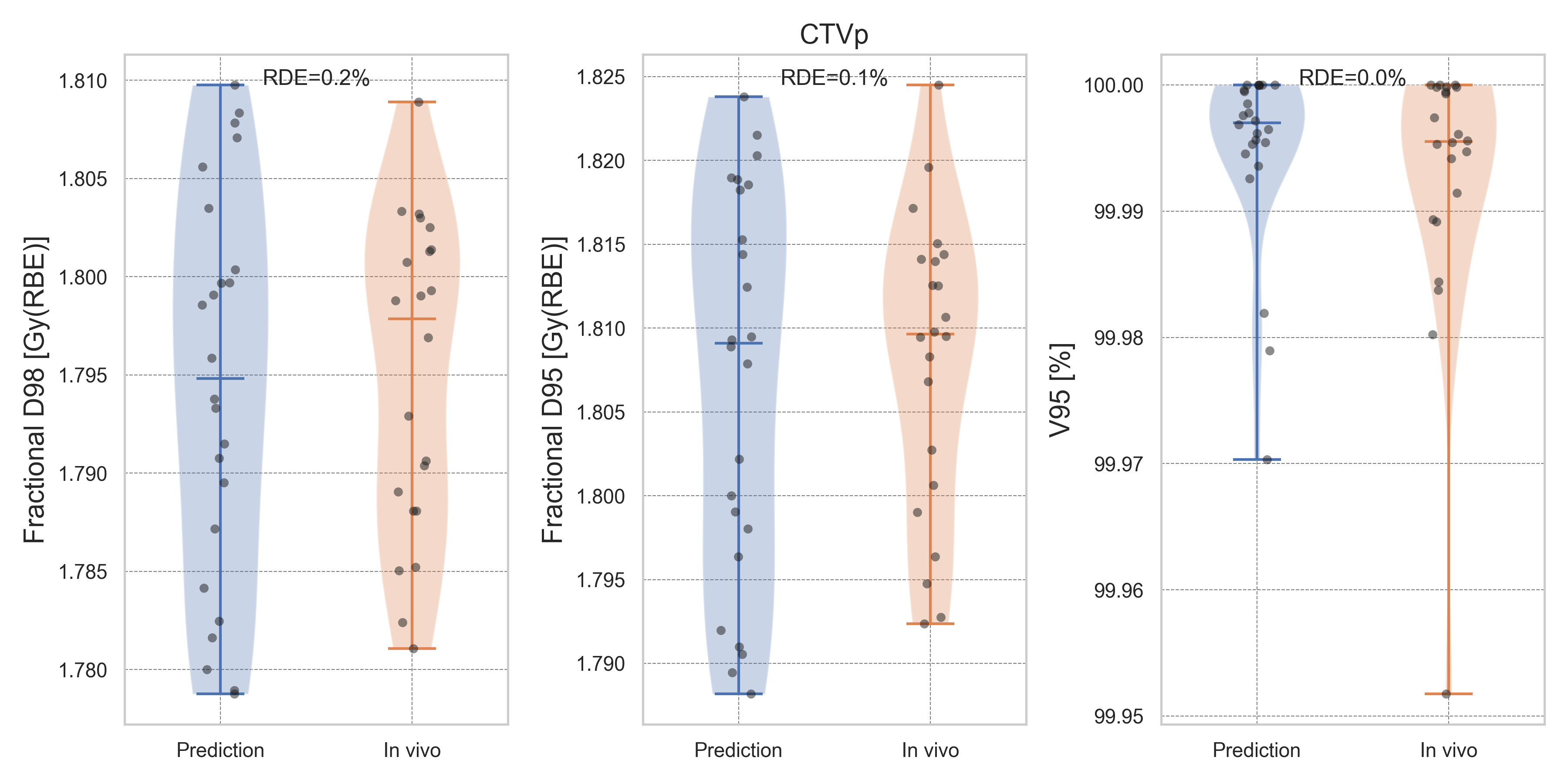}
        \caption{}
        \label{fig:subfig2}
    \end{subfigure}
    \caption{Violin plots of the D98, D95, and V95 distributions for the predictive and \textit{in vivo} methods computed with FRED Monte Carlo dose engine. (a) Lung cancer patients (CTVp and CTVn); (b) Esophageal cancer patients (CTVp only).}
    \label{fig:Aprior_Apost_fx}
\end{figure}

The predictive method, based on 24 synthetic sinusoidal breathing scenarios, was compared against the \textit{in vivo} model that integrates patient-specific breathing traces and machine log files. Figure~\ref{fig:Aprior_Apost_fx} shows the fractional DVH distributions (D98, D95, and V95) for both lung (panel a) and esophageal (panel b) cancer patients. The violin plots demonstrate that the predictive method accurately reproduces the dosimetric variability observed in the \textit{in vivo} calculations. Quantitatively, the RDE between the two approaches was below 1\% for both CTVp and CTVn, confirming strong concordance.

\begin{figure}[htbp]
    \centering
    \begin{subfigure}[b]{0.7\textwidth}
        \centering
        \includegraphics[width=\linewidth]{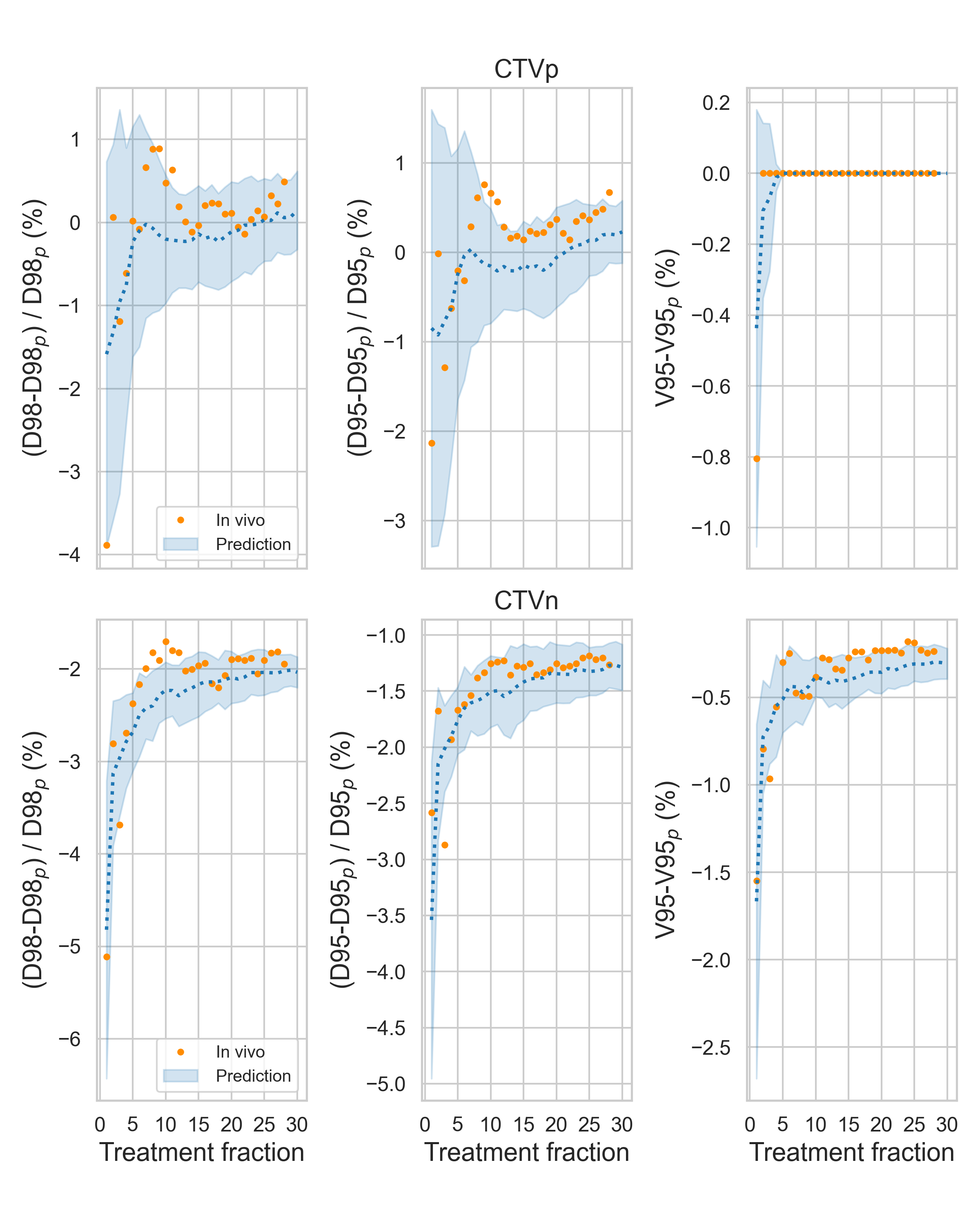}
        \caption{}
    \end{subfigure}
    \hfill
    \begin{subfigure}[b]{0.7\textwidth}
        \centering
        \includegraphics[width=\linewidth]{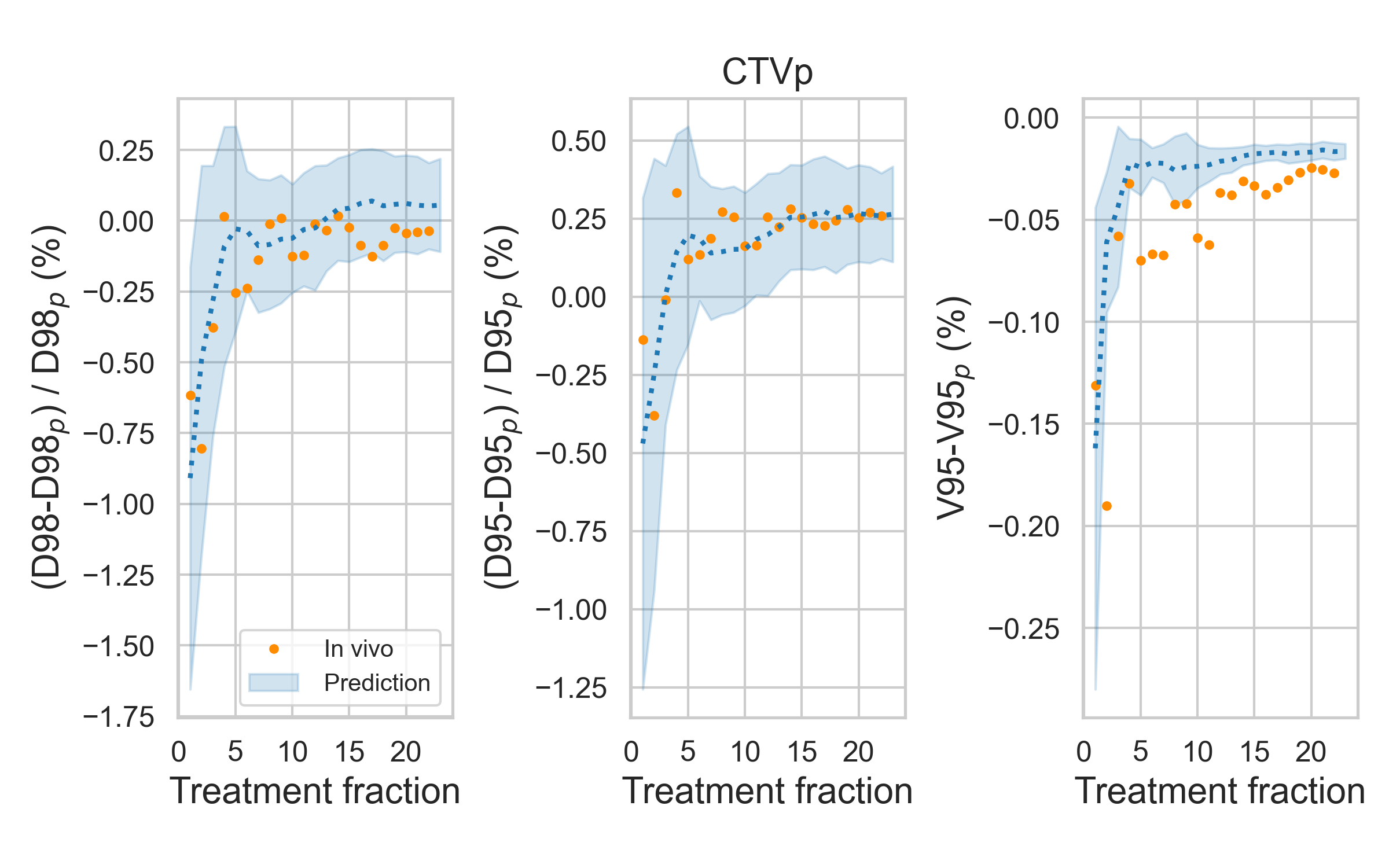}
        \caption{}
    \end{subfigure}
    \caption{Cumulative DVH parameters across the treatment course for the predictive and \textit{in vivo} approaches. (a) Lung cancer patient (CTVp and CTVn); (b) Esophageal cancer patient (CTVp only).}
    \label{fig:Aprior_Apost_cum}
\end{figure}

To investigate the mitigation of interplay effects through fractionation, we analyzed cumulative DVH metrics across the treatment course (Figure~\ref{fig:Aprior_Apost_cum}). In these plots, the predictive model is represented by a shaded band (indicating standard deviation from 10 random fractionation schemes) and a dashed line (mean value). The \textit{in vivo} results are shown as orange circles.

In the lung cancer case (Figure~\ref{fig:Aprior_Apost_cum}(a)), the cumulative DVH metrics for the primary tumor (CTVp) converged to within 2\% of the nominal plan after just five fractions, indicating that the interplay effect was effectively averaged out early in the treatment course. A similar trend was observed for the nodal volume (CTVn), although the cumulative dose remained slightly lower than planned throughout the treatment. At the end of the course, the D98 and D95 for CTVn were 2.0\%~$\pm$~0.2\% and 1.3\%~$\pm$~0.2\% below the nominal plan, respectively. It is important to note that, following our current clinical practice, all nodal regions are treated as a single structure (CTVn), and no motion management or volume expansion—such as an internal target volume (ITV) is applied to CTVn~\cite{taasti2025proton}.

In the esophageal cancer case (Figure~\ref{fig:Aprior_Apost_cum}(b)), both the predictive and \textit{in vivo} methods produced cumulative DVH values that remained within 2\% of the planned dose starting from the first fraction. This close agreement across all fractions highlights the robustness of the treatment plan under free-breathing conditions, even in the presence of respiratory motion.

\section{Discussion}
This study presents a comprehensive framework for predicting interplay effects combined with range and setup uncertainties for a Mevion Hyperscan S250i PBS proton therapy system. The focus is on lung and esophageal cancer patients treated under free-breathing conditions. We proposed a predictive interplay assessment method based on synthetic breathing traces and compared it to an \textit{in vivo} approach that incorporates patient-specific respiratory traces and machine log files. Both methods were implemented within the RayStation treatment planning system and validated using the independent Monte Carlo dose engine FRED.

The two dose engines demonstrated excellent agreement, with mean dose differences below 1\%, which is on the same order as the calculation uncertainty of the clinical TPS, thereby confirming the robustness of the implementation (Figure~\ref{fig:FREDVsTPS}). Thanks to its high computational speed and exceptional versatility, FRED provides a powerful platform for advanced research and rapid development of novel interplay evaluation workflows. In parallel, the TPS—with its integrated graphical user interface—remains the more user-friendly environment for routine clinical use.

Application of the two approaches to clinical cases yielded highly consistent results, with relative differences below 2\% across all evaluated DVH metrics (Figure~\ref{fig:Aprior_Apost_fx}). Although the predictive model assumes a sinusoidal breathing pattern and omits inter-fractional delivery variations, its strong agreement with the \textit{in vivo} evaluation suggests that these simplifications have minimal impact on clinical interplay estimation \cite{pastor2021should}. Furthermore, the 24 breathing scenarios used in the predictive model effectively captured the range of \textit{in vivo} dose variations.

The convergence of cumulative dose distributions toward the nominal plan (Figure~\ref{fig:Aprior_Apost_cum}) further confirms the well-established mitigation of interplay effects through fractionation \cite{seco2009breathing, grassberger2015motion}. For the primary CTV (CTVp), only minor deviations from the nominal dose were observed, even when accounting for respiratory motion, range uncertainties, and setup deviations. DVH parameters converged to within 2\% of planned values after just five fractions, demonstrating plan robustness under conventional fractionation. These findings align with prior work by Meijers et al. \cite{meijers2020evaluation}, who also found that organ motion has a relatively minor impact on dose degradation compared to anatomical changes—though their study employed rescanning. 

Our results suggest that the relatively large spot sizes of the Mevion Hyperscan S250i system inherently mitigate interplay effects, eliminating the need for time-consuming rescanning techniques. This observation is consistent with previous works \cite{rana2022small, dowdell2013interplay, grewal2023dosimetric} and has been further substantiated by phantom studies conducted at our institution.

In the lung cancer case, a comparative analysis between the CTVp and CTVn revealed greater robustness for the former. The CTVp was optimized using 3D robust optimization based on the ITV, further expanded by 1 mm, whereas the CTVn was optimized without any additional margins, resulting in reduced dosimetric robustness. This confirms that the ITV robust optimization improves the robustness also towards organ motion and interplay \cite{inoue2016limited}.
Despite some loss of coverage in the CTVp during the first five fractions, and a more pronounced degradation in the CTVn over the full treatment course, the biological implications may be limited, given the small magnitude of these deviations \cite{bortfeld2006biologic}.

\subsection*{Limitations}
A primary limitation of this study is the exclusion of inter-fractional anatomical changes and variations in tumor motion amplitude throughout the treatment course. Interplay effects were assessed using the planning 4DCT for each fraction, which may limit generalization, particularly in patients experiencing significant anatomical evolution or respiratory variability. While DIR enables dose accumulation across repeated CTs, it introduces uncertainties, particularly in the presence of substantial geometric changes. Several studies, including Ribeiro et al.~\cite{ribeiro2018assessment} and Stützer et al.~\cite{stutzer2016evaluation}, have highlighted both the capabilities and limitations of DIR in proton therapy. Before integrating accumulated dose evaluations into the predictive framework, DIR errors must be rigorously quantified. A recent study by Taasti et al. \cite{taasti2025dose} introduced a method for dose accumulation that accounts for setup and range uncertainties. This approach considers anatomical changes between fractions and also facilitates the QA of deformable dose mapping.

\begin{figure}
    \centering
    \includegraphics[width=1\linewidth]{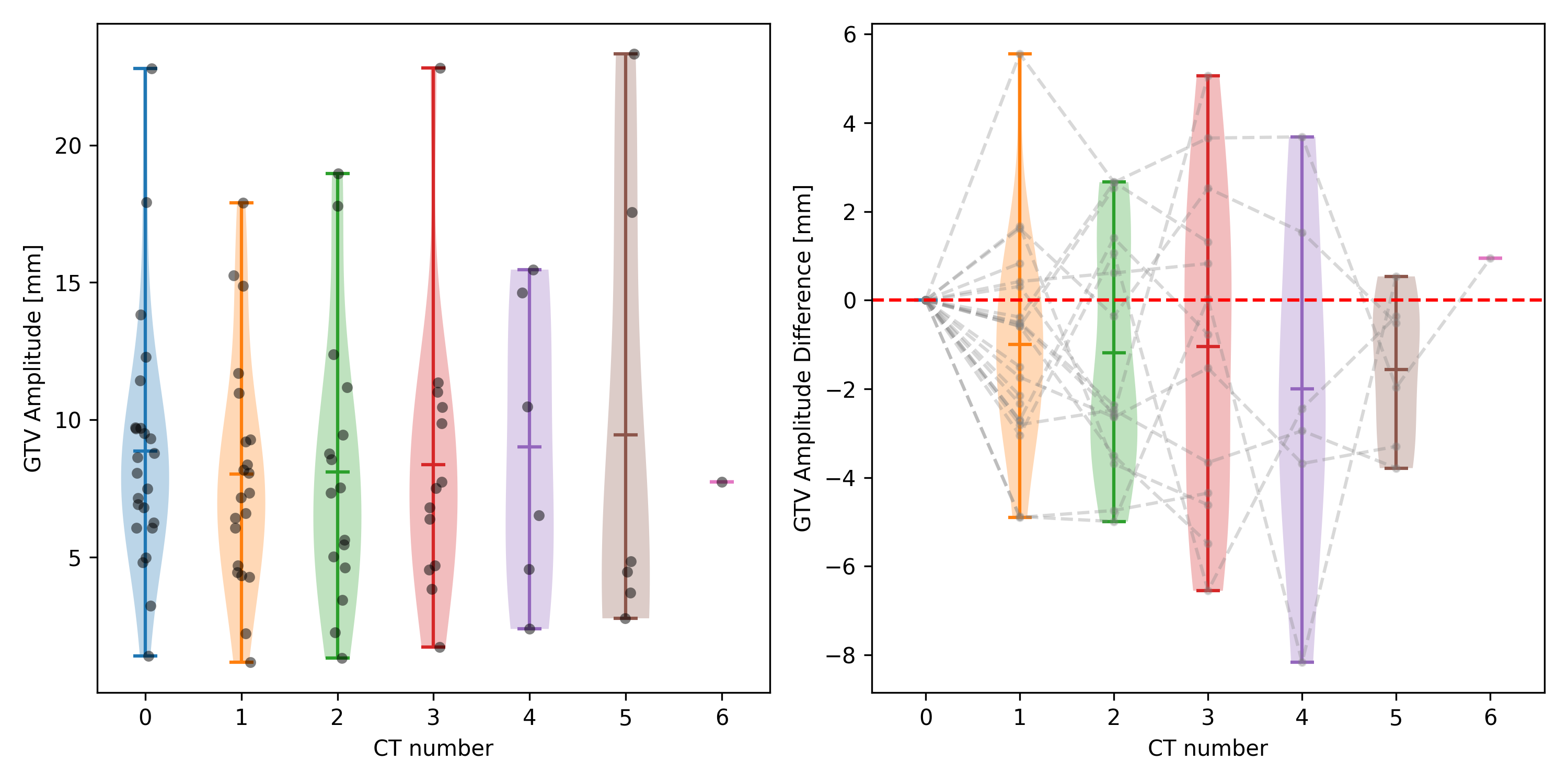}
    \caption{Left panel: distribution of the GTV amplitude as a function of repeated 4DCT number. The right panel shows the difference between the amplitude in the repeated CT and the first planning CT (CT number 0).}
    \label{fig:TumorAmplitude}
\end{figure}

On the other hand, data extracted from our lung patient cohort indicate a general trend of decreasing tumor motion amplitude over the course of treatment, supporting the assumption that the planning 4DCT remains a reasonable surrogate for respiratory motion modeling. Figure~\ref{fig:TumorAmplitude} shows the GTV amplitude of 24 patients treated at our institution from mid-2024 onward, computed as the maximum 1D displacement of the GTV center of mass over the three orthogonal directions. This trend was also confirmed by a previous study analyzing patients treated from 2019 to 2023~\cite{taasti2025proton}. In cases requiring adaptive planning due to anatomical or motion changes, the predictive model can be reapplied using the updated treatment plan and planning 4DCT, which intrinsically accounts for the new tumor amplitude. Notably, the predictive evaluation described here is intended for pre-treatment use, when no information on future anatomical changes is available.

Another limitation relates to the \textit{in vivo} model’s assumption of phase alignment between the patient-specific breathing trace and the 4DCT phase at the start of treatment. This simplification can introduce uncertainties in interplay calculations, as shown in previous studies \cite{dowdell2013interplay}. However, the predictive model incorporates all initial phase scenarios across 4DCT phases, inherently addressing this uncertainty. The strong agreement between the predictive and \textit{in vivo} models supports the validity of this assumption in practice.

\section{Conclusions}
In this study, we presented a predictive approach to evaluate the robustness of proton therapy plans against organ motion, range, and setup uncertainties before the start of treatment. The model was validated using \textit{in vivo} data incorporating patient- and fraction-specific parameters, such as breathing signals and machine log files. Two clinical cases, one lung and one esophageal cancer patient treated under free-breathing conditions without rescanning, demonstrated only minor dose degradation in the primary CTV due to organ motion. 

While these results are encouraging, further investigation in a larger patient cohort is necessary to confirm these preliminary findings. This work is currently ongoing and will be presented in a separate publication. If validated, the proposed predictive framework could support clinical decision-making and enable the adoption of hypofractionated treatment regimens, as well as the treatment of lung targets exhibiting large respiratory motion (i.e., ~$>$~2cm).

%\section*{Supplementary Materials}
%\renewcommand{\thefigure}{S\arabic{figure}}
%\setcounter{figure}{0}

%\subsection*{Tumor amplitude variations over the treatment course}
%Figure~\ref{fig: Tumor Amplitude} shows the GTV amplitude of the patients treated in our institution from mid-2024 to now. The amplitude was calculated as the maximum 1D displacement of the GTV center of mass over the three directions.
%This was also confirmed by a previous study where the GTV amplitudes were calculated on the patient treated from 2019 to 2023 \cite{taasti2025proton}.
%In addition, in case of plan adaptation, we use the new planning 4DCT for our interplay evaluation which intrinsically takes into account the new tumor amplitude.

\section*{Acknowledgments}
\section*{References}
\bibliographystyle{apalike}
%\bibliography{References}
\bibliography{main}

\begin{thebibliography}{}

\bibitem[Bengtsson et~al., 2025]{bengtsson2025interplay}
Bengtsson, I., Forsgren, A., Fredriksson, A., and Zhang, Y. (2025).
\newblock Interplay-robust optimization for treating irregularly breathing lung patients with pencil beam scanning.
\newblock {\em Medical Physics}.

\bibitem[Bert and Durante, 2011]{bert2011motion}
Bert, C. and Durante, M. (2011).
\newblock Motion in radiotherapy: particle therapy.
\newblock {\em Physics in Medicine \& Biology}, 56(16):R113.

\bibitem[Bortfeld and Paganetti, 2006]{bortfeld2006biologic}
Bortfeld, T. and Paganetti, H. (2006).
\newblock The biologic relevance of daily dose variations in adaptive treatment planning.
\newblock {\em International Journal of Radiation Oncology* Biology* Physics}, 65(3):899--906.

\bibitem[Canters et~al., 2024]{canters2024robustness}
Canters, R., van~der Klugt, K., Taasti, V.~T., Buijsen, J., Ta, B., Steenbakkers, I., Houben, R., Vilches-Freixas, G., and Berbee, M. (2024).
\newblock Robustness of intensity modulated proton treatment of esophageal cancer for anatomical changes and breathing motion.
\newblock {\em Radiotherapy and Oncology}, 198:110409.

\bibitem[Chang et~al., 2014]{chang2014clinical}
Chang, J.~Y., Li, H., Zhu, X.~R., Liao, Z., Zhao, L., Liu, A., Li, Y., Sahoo, N., Poenisch, F., Gomez, D.~R., et~al. (2014).
\newblock Clinical implementation of intensity modulated proton therapy for thoracic malignancies.
\newblock {\em International Journal of Radiation Oncology* Biology* Physics}, 90(4):809--818.

\bibitem[Chang et~al., 2017]{chang2017consensus}
Chang, J.~Y., Zhang, X., Knopf, A., Li, H., Mori, S., Dong, L., Lu, H.-M., Liu, W., Badiyan, S.~N., Both, S., et~al. (2017).
\newblock Consensus guidelines for implementing pencil-beam scanning proton therapy for thoracic malignancies on behalf of the ptcog thoracic and lymphoma subcommittee.
\newblock {\em International Journal of Radiation Oncology* Biology* Physics}, 99(1):41--50.

\bibitem[Dowdell et~al., 2013]{dowdell2013interplay}
Dowdell, S., Grassberger, C., Sharp, G., and Paganetti, H. (2013).
\newblock Interplay effects in proton scanning for lung: a 4d monte carlo study assessing the impact of tumor and beam delivery parameters.
\newblock {\em Physics in Medicine \& Biology}, 58(12):4137.

\bibitem[Engwall et~al., 2018]{engwall20184d}
Engwall, E., Fredriksson, A., and Glimelius, L. (2018).
\newblock 4d robust optimization including uncertainties in time structures can reduce the interplay effect in proton pencil beam scanning radiation therapy.
\newblock {\em Medical physics}, 45(9):4020--4029.

\bibitem[Gajewski et~al., 2020]{gajewski2020implementation}
Gajewski, J., Schiavi, A., Krah, N., Vilches-Freixas, G., Rucinski, A., Patera, V., and Rinaldi, I. (2020).
\newblock Implementation of a compact spot-scanning proton therapy system in a gpu monte carlo code to support clinical routine.
\newblock {\em Frontiers in Physics}, 8:578605.

\bibitem[Grassberger et~al., 2015]{grassberger2015motion}
Grassberger, C., Dowdell, S., Sharp, G., and Paganetti, H. (2015).
\newblock Motion mitigation for lung cancer patients treated with active scanning proton therapy.
\newblock {\em Medical physics}, 42(5):2462--2469.

\bibitem[Grewal et~al., 2023]{grewal2023dosimetric}
Grewal, H.~S., Ahmad, S., and Jin, H. (2023).
\newblock Dosimetric study of the interplay effect using three-dimensional motion phantom in proton pencil beam scanning treatment of moving thoracic tumours.
\newblock {\em Journal of Radiotherapy in Practice}, 22:e11.

\bibitem[Inoue et~al., 2016]{inoue2016limited}
Inoue, T., Widder, J., van Dijk, L.~V., Takegawa, H., Koizumi, M., Takashina, M., Usui, K., Kurokawa, C., Sugimoto, S., Saito, A.~I., et~al. (2016).
\newblock Limited impact of setup and range uncertainties, breathing motion, and interplay effects in robustly optimized intensity modulated proton therapy for stage iii non-small cell lung cancer.
\newblock {\em International Journal of Radiation Oncology* Biology* Physics}, 96(3):661--669.

\bibitem[Knopf et~al., 2011]{knopf2011scanned}
Knopf, A.-C., Hong, T.~S., and Lomax, A. (2011).
\newblock Scanned proton radiotherapy for mobile targets—the effectiveness of re-scanning in the context of different treatment planning approaches and for different motion characteristics.
\newblock {\em Physics in Medicine \& Biology}, 56(22):7257.

\bibitem[Li et~al., 2006]{li2006technical}
Li, X.~A., Stepaniak, C., and Gore, E. (2006).
\newblock Technical and dosimetric aspects of respiratory gating using a pressure-sensor motion monitoring system.
\newblock {\em Medical physics}, 33(1):145--154.

\bibitem[Li et~al., 2014]{li2014interplay}
Li, Y., Kardar, L., Li, X., Li, H., Cao, W., Chang, J.~Y., Liao, L., Zhu, R.~X., Sahoo, N., Gillin, M., et~al. (2014).
\newblock On the interplay effects with proton scanning beams in stage iii lung cancer.
\newblock {\em Medical physics}, 41(2):021721.

\bibitem[Lu et~al., 2006]{lu2006comparison}
Lu, W., Parikh, P.~J., Hubenschmidt, J.~P., Bradley, J.~D., and Low, D.~A. (2006).
\newblock A comparison between amplitude sorting and phase-angle sorting using external respiratory measurement for 4d ct.
\newblock {\em Medical physics}, 33(8):2964--2974.

\bibitem[Meijers et~al., 2020]{meijers2020evaluation}
Meijers, A., Knopf, A.-C., Crijns, A.~P., Ubbels, J.~F., Niezink, A.~G., Langendijk, J.~A., Wijsman, R., and Both, S. (2020).
\newblock Evaluation of interplay and organ motion effects by means of 4d dose reconstruction and accumulation.
\newblock {\em Radiotherapy and Oncology}, 150:268--274.

\bibitem[Pastor-Serrano et~al., 2021]{pastor2021should}
Pastor-Serrano, O., Habraken, S., Lathouwers, D., Hoogeman, M., Schaart, D., and Perk{\'o}, Z. (2021).
\newblock How should we model and evaluate breathing interplay effects in impt?
\newblock {\em Physics in Medicine \& Biology}, 66(23):235003.

\bibitem[Rana and Rosenfeld, 2021]{rana2021investigating}
Rana, S. and Rosenfeld, A.~B. (2021).
\newblock Investigating volumetric repainting to mitigate interplay effect on 4d robustly optimized lung cancer plans in pencil beam scanning proton therapy.
\newblock {\em Journal of Applied Clinical Medical Physics}, 22(3):107--118.

\bibitem[Rana and Rosenfeld, 2022]{rana2022small}
Rana, S. and Rosenfeld, A.~B. (2022).
\newblock Small spot size versus large spot size: Effect on plan quality for lung cancer in pencil beam scanning proton therapy.
\newblock {\em Journal of Applied Clinical Medical Physics}, 23(2):e13512.

\bibitem[Ribeiro et~al., 2018]{ribeiro2018assessment}
Ribeiro, C.~O., Knopf, A., Langendijk, J.~A., Weber, D.~C., Lomax, A.~J., and Zhang, Y. (2018).
\newblock Assessment of dosimetric errors induced by deformable image registration methods in 4d pencil beam scanned proton treatment planning for liver tumours.
\newblock {\em Radiotherapy and Oncology}, 128(1):174--181.

\bibitem[Schiavi et~al., 2017]{schiavi2017fred}
Schiavi, A., Senzacqua, M., Pioli, S., Mairani, A., Magro, G., Molinelli, S., Ciocca, M., Battistoni, G., and Patera, V. (2017).
\newblock Fred: a gpu-accelerated fast-monte carlo code for rapid treatment plan recalculation in ion beam therapy.
\newblock {\em Physics in Medicine \& Biology}, 62(18):7482.

\bibitem[Seco et~al., 2009]{seco2009breathing}
Seco, J., Robertson, D., Trofimov, A., and Paganetti, H. (2009).
\newblock Breathing interplay effects during proton beam scanning: simulation and statistical analysis.
\newblock {\em Physics in Medicine \& Biology}, 54(14):N283.

\bibitem[Spautz et~al., 2023]{spautz2023comparison}
Spautz, S., Haase, L., Tschiche, M., Makocki, S., Richter, C., Troost, E.~G., and St{\"u}tzer, K. (2023).
\newblock Comparison of 3d and 4d robustly optimized proton treatment plans for non-small cell lung cancer patients with tumour motion amplitudes larger than 5 mm.
\newblock {\em Physics and Imaging in Radiation Oncology}, 27:100465.

\bibitem[St{\"u}tzer et~al., 2016]{stutzer2016evaluation}
St{\"u}tzer, K., Haase, R., Lohaus, F., Barczyk, S., Exner, F., L{\"o}ck, S., R{\"u}haak, J., Lassen-Schmidt, B., Corr, D., and Richter, C. (2016).
\newblock Evaluation of a deformable registration algorithm for subsequent lung computed tomography imaging during radiochemotherapy.
\newblock {\em Medical physics}, 43(9):5028--5039.

\bibitem[Taasti et~al., 2025a]{taasti2025dose}
Taasti, V.~T., Hattu, D., Berb{\'e}e, M., de~Ruysscher, D., and Canters, R. (2025a).
\newblock Dose accumulation over the treatment course including setup and range uncertainty for proton therapy of esophageal and lung cancer patients.
\newblock {\em Medical Physics}, 52(8):e18037.

\bibitem[Taasti et~al., 2021]{taasti2021treatment}
Taasti, V.~T., Hattu, D., Vaassen, F., Canters, R., Velders, M., Mannens, J., van Loon, J., Rinaldi, I., Unipan, M., and van Elmpt, W. (2021).
\newblock Treatment planning and 4d robust evaluation strategy for proton therapy of lung tumors with large motion amplitude.
\newblock {\em Medical Physics}, 48(8):4425--4437.

\bibitem[Taasti et~al., 2025b]{taasti2025proton}
Taasti, V.~T., Kneepkens, E., van~der Stoep, J., Velders, M., Cobben, M., Vullings, A., Buck, J., Visser, F., van~den Bosch, M., Hattu, D., et~al. (2025b).
\newblock Proton therapy of lung cancer patients--treatment strategies and clinical experience from a medical physicist’s perspective.
\newblock {\em Physica Medica}, 130:104890.

\bibitem[Vilches-Freixas et~al., 2020]{vilches2020beam}
Vilches-Freixas, G., Unipan, M., Rinaldi, I., Martens, J., Roijen, E., Almeida, I.~P., Decabooter, E., and Bosmans, G. (2020).
\newblock Beam commissioning of the first compact proton therapy system with spot scanning and dynamic field collimation.
\newblock {\em The British journal of radiology}, 93(1107):20190598.

\bibitem[Walter et~al., 2016]{walter2016evaluation}
Walter, F., Freislederer, P., Belka, C., Heinz, C., S{\"o}hn, M., and Roeder, F. (2016).
\newblock Evaluation of daily patient positioning for radiotherapy with a commercial 3d surface-imaging system (catalyst™).
\newblock {\em Radiation oncology}, 11(1):154.

\end{thebibliography}
\end{document}